\documentclass[submission, Phys]{SciPost}
\hypersetup{
    colorlinks,
    linkcolor={red!50!black},
    citecolor={blue!50!black},
    urlcolor={blue!80!black}
}

\usepackage[bitstream-charter]{mathdesign}
\DeclareSymbolFont{usualmathcal}{OMS}{cmsy}{m}{n}
\DeclareSymbolFontAlphabet{\mathcal}{usualmathcal}

\usepackage[normalem]{ulem} % for crossing out sentences to highlight changes to the draft
\graphicspath{{./img/},{./img/tex/}}

\begin{document}

\begin{center}{\Large \textbf{
Thermal pure matrix product state in two dimensions: tracking thermal equilibrium 
from paramagnet down to the Kitaev honeycomb spin liquid state\\
}}\end{center}

% TODO: write the author list here. Use first name (+ other initials) + surname format.
% Separate subsequent authors by a comma, omit comma and use "and" for the last author.
% Mark the corresponding author with a superscript star.
\begin{center}
Matthias Gohlke\textsuperscript{1},
Atsushi Iwaki\textsuperscript{2} and
Chisa Hotta\textsuperscript{2}
\end{center}

\begin{center}
{\bf 1} Theory of Quantum Matter Unit, Okinawa Institute of Science and Technology Graduate University, Onna-son, Okinawa 904-0495, Japan
\\
{\bf 2} Department of Basic Science, The University of Tokyo, Meguro-ku, Tokyo 153-8902, Japan
and Komaba Institute for Science, The University of Tokyo, Meguro-ku, Tokyo 153-8902, Japan
\\
% ${}^\star$ {\small \sf CorrespondingAuthor@email.address}
\end{center}

\begin{center}
\today
\end{center}

% For convenience during refereeing (optional),
% you can turn on line numbers by uncommenting the next line:
%\linenumbers
\section*{Abstract}
{\bf
% TODO: write your abstract here.
We present the first successful application of the matrix product state (MPS) representing a thermal quantum pure state (TPQ) 
in equilibrium in two spatial dimensions over almost the entire temperature range. 
We use the Kitaev honeycomb model as a prominent example hosting a quantum spin liquid (QSL) ground state 
to target the two specific-heat peaks previously solved nearly exactly using the free Majorana fermionic description. 
Starting from the high-temperature random state, our TPQ-MPS framework on a cylinder precisely reproduces these peaks, 
showing that the quantum many-body description based on spins can still capture 
the emergent itinerant Majorana fermions in a ${\mathbb Z}_2$ gauge field. 
The truncation process efficiently discards the high-energy states, 
eventually reaching the long-range entangled topological state 
approaching the exact ground state for a given finite size cluster.
An advantage of TPQ-MPS over exact diagonalization or purification-based methods is its lowered numerical cost coming from a reduced effective Hilbert space even at finite temperature.
}

\vspace{10pt}
\noindent\rule{\textwidth}{1pt}
\tableofcontents\thispagestyle{fancy}
\noindent\rule{\textwidth}{1pt}
\vspace{10pt}

\section{Introduction}
\label{sec:intro}
\begin{figure}[tb!]
    \centering
    \includegraphics[scale=0.85]{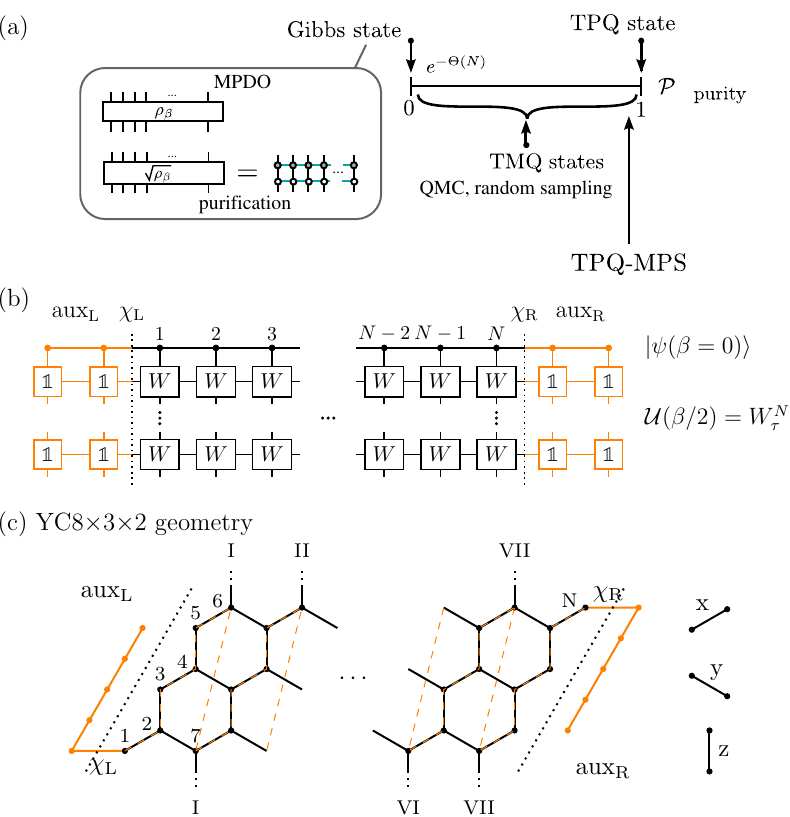} 
    \caption{(a) Schematic illustration of different thermal quantum states, classified by purity. 
              The two well-known representations of the Gibbs state are sketched. 
              Purification prepares a product state of singlets on a pair of system and ancilla sites, 
              and operates $\sqrt{\rho_B}\otimes U_{\rm aux}$ where $\sqrt{\rho_B}=e^{-\beta \hat H/2}$ acts on the system
              and $U_{\rm aux}$ on the ancilla. The resultant state is shown in the simplified tensor form. 
              (b) Schematic illustration of TPQ-MPS with auxiliary degrees of freedom providing an entanglement bath, and the MPO-based imaginary time evolution. 
             (c) Illustration of the honeycomb lattice composed of $x$-, $y$-, and $z$- bonds. 
             For the underlying 1D MPS structure, we use a cylindrical geometry with a helical enumeration scheme
             as is highlighted by orange dashed lines. 
             Equivalent bonds across the boundary are marked with the same roman literal.
             Specifically shown is the YC8$\times$3$\times$2 geometry with a shifted (by one lattice vector) boundary condition.
             Orange dots connected by solid lines represent the auxiliary sites. 
             }
    \label{f1}
\end{figure}

% TODO: write your article here.
Characterizing a thermal quantum state, a quantum many-body state at finite temperature  
is an ongoing fundamental challenge in condensed matter physics and beyond, 
since it is often a matter of quantum and classical correlations 
studied in statistical and quantum information physics\cite{Nielsen2000}. 
Such a state has an intriguing aspect 
in that its representation is largely left facultative\cite{Iwaki2022}; 
the Gibbs state is a mixture of an exponential number of states 
given by the density matrix $\rho_\beta$ of small purity 
$\mathcal P \sim e^{-\Theta(N)}$, i.e., vanishing exponentially with the system size $N$. 
The thermal pure quantum (TPQ) state, on the other hand, is a single pure state of purity $\mathcal P = 1$. 
In addition, there exist numerous thermal mixed quantum (TMQ) states with a purity between Gibbs and TPQ (see Fig.~\ref{f1}(a)). 
Canonical typicality guarantees that all these choices equivalently yield 
the same thermal equilibrium properties of the subsystem\cite{Popescu2006, Goldstein2006}, 
and are macroscopically in the ``same" thermal state. 
Since Gibbs, TPQ, and TMQ states rely on different design concepts, 
even when applying the ``same" tensor network representation, 
its structure, convergence, or the amount of numerical resources required
likely depend on which type of thermal state is chosen.
%*%*%*%*%*%*%*%*%*%*%*%*%*%*%*%*%*%*%*%*%*%*%*%*%*%*%*
%*%*%*%*%*%*%*%*%*%*%*%*%*%*%*%*%*%*%*%*%*%*%*%*%*%*%*
\par
An important development concerning the Gibbs state is the matrix product density operator (MPDO), 
which provides a direct tensor network representation of the density matrix operator, $\rho_\beta$ \cite{Verstraete2004, Zwolak2004}. 
Another standard form of the Gibbs state is the purified state analog to thermofield double,
consisting of the size-$N$ system and the same numbers of ancilla degrees of freedom
each suspended to a local site\cite{Feiguin2005}. 
Ancilla serve as an entanglement bath and tracing out the ancilla corresponds to taking the Gibbs ensemble. 
These \emph{doubled} states also conform to a matrix product operator (MPO) approach 
%%%%%%
\footnote{
The difference between MPDO and purification is that 
the MPDO is not necessarily positive definite after truncation, 
whereas purification using a canonical form is positive definite. 
However, purification generally requires larger $\chi$ than MPDO \cite{Cuevas13}, 
and there are some examples\cite{Goto2021} that 
the purification MPO shows a divergence of $\chi$ at low temperatures, 
which may indicate that the thermal area law may not safely apply. 
}, 
whose schematic illustrations are shown in Fig.~\ref{f1}(a).
%%%%%%
Here, the entanglement entropy is meaningless as a measure to characterize the Gibbs state.
Instead, the thermal area law of mutual information between subsystems determines
the bond dimension $\chi$ of MPO's\cite{Wolf2008,Barthel2017,Kuwahara2021}. 
The numerical drawback of MPDO or purification is the 
increase of the Hilbert space dimension due to the doubled degrees of freedom.
Still, MPDO has been developed further recently using the XTRG algorithm\cite{Chen2018}, 
which realizes an exponential cooling down of the system by 
iteratively multiplying the matrix $\rho_\beta \times \rho_\beta=\rho_{2\beta}$, 
allowing to reach very low temperatures rapidly. 
XTRG has successfully been applied to two dimensions including our target\cite{Li2019,Li2020}, the Kitaev honeycomb model\cite{Kitaev2006}.
\par
The TPQ state, in comparison, consisting only of physical degrees of freedom, is pure by construction, 
and does not need the doubling of the local Hilbert space. 
In MPDO and its analogues, the doubling or the ancilla play the role of an ensemble average---%
or the classical mixture of states---which provide the volume-law thermal entropy. 
The lack of doubling implies that the pure TPQ state needs to store the same amount of entropy 
internally as a volume-law entanglement entropy\cite{Garrison2018, Nakagawa2018, Iwaki2021}. 
For such purpose, the tensor-network-based representation bounded by the area law entanglement 
are thought to naturally be out of reach. 
Yet, the authors have recently exploited the specific form of matrix product state (MPS) 
practically recovering the volume law entanglement; 
only two ancilla/auxiliaries are attached to both edges of the one-dimensional (1D) MPS train, 
yet they have turned out to be sufficient to keep the nearly uniform distribution of entanglement entropy density 
throughout the system
%%%%%
\footnote{If we take a bipartition of the TPQ-MPS system into left and right, each attached to the auxiliary, 
the entanglement entropy does not depend on the size of the left/right part, 
unlike the usual MPS that follows the size-dependent Page curve. 
This translational invariance of the entanglement entropy 
allows entanglement entropy between the center-$n$ sites and the rest (with $N-n$ sites and two auxiliaries) to follow 
the $n$-linear volume law (see Ref.~\cite{Iwaki2021}). 
}
%%%%
which is essential for the volume law entanglement.
We call this construction the TPQ-MPS\cite{Iwaki2021}.
The TPQ state itself has a numerically long history\cite{Imada1986,Jaklic1994,Hams2000,Iitaka2004} 
far before the formulative seminal works\cite{Sugiura2012, Sugiura2013}. 
They mostly rely on a full Hilbert space representation 
using Lanczos-based methods that limit the system size to typically $N\lesssim 30-40$. 
The TPQ-MPS largely shrinks the representation space and increases $N$ by factors 
by efficiently choosing its constituent states to those representing 
the target temperature limited by the bond dimension of the MPS. 
We review a measure of the quality of a TPQ-MPS, which has been developed in Ref.~\cite{Iwaki2022}, in Appendix A.
\par
The present work advances a few steps in developing a TPQ-MPS for two dimensions (2D), 
particularly for a quantum mechanically nontrivial quantum spin liquid state with long-range entanglement. 
Encoding the substantial amount of entanglement expected for QSL within an MPS or a tensor-network 
is generically a challenging task, although reported in the case of ground state\cite{Xiang2017,Yin-Chen2017,Gohlke2017}. 
Our result is the first to track the state by an MPS 
in the nearly pure form from the high-temperature random state 
down to the QSL with substantial entanglement 
between limited selection of basis states. 
\par
We finally refer to some TMQ-state-based approaches; 
the minimally entangled typical thermal state (METTS)\cite{White2009, Stoudenmire2010} 
mixes (takes an equal weight average of) a series of MPS generated from the Markov process. 
The quantum Monte Carlo designs a local product state basis to suppress the sign problem 
\cite{Mila2021,Wietek2019}, which are recently highlighted in combination with the iPEPS. 
\par
%%%%%%%%%%%%%%%%%%%%%%%%%%%%%%%%%%%%%%%%%%%%%%%%%%%%%%%%%%%%%%%%%%%%%%%%%%%%%%%
%%%%%%%%%%%%%%%%%%%%%%%%%%%%%%%%%%%%%%%%%%%%%%%%%%%%%%%%%%%%%%%%%%%%%%%%%%%%%%%
%%%%%%%%%%%%%%%%%%%%%%%%%%%%%%%%%%%%%%%%%%%%%%%%%%%%%%%%%%%%%%%%%%%%%%%%%%%%%%%
%\\
\section{Construction of the TPQ-MPS state}
We consider the standard imaginary-time evolution in generating the TPQ state
at inverse temperature $\beta = 1/T$ given as
\begin{equation}
   |\Psi_\beta \rangle = e^{-\frac{\beta}{2}\mathcal{H}} |\Psi_0\rangle\; =
     \mathcal U(\beta/2)\:  |\Psi_0\rangle ,
\label{eqn:TPQ-state}
\end{equation}
where $\mathcal H$ is the Hamiltonian of the system of interest,
and the initial state $|\Psi_0\rangle$ representing an 'infinite-$T$' state is chosen as random, 
satisfying $\overline{|\Psi_0\rangle\langle \Psi_0|}\propto I$, 
where $\overline{\cdots}$ is the random average and 
$I$ is the unit matrix.
\par
We now specify the construction of TPQ-MPS utilized here.
The 1D tensor train of size-$N$ and bond dimension $\chi$ is prepared with auxiliary degrees of freedom added to both ends
to provide an entanglement bath (Fig.~\ref{f1}(b)).
Here, instead of the $\chi\times\chi$ form proposed in Ref.\cite{Iwaki2021},
each auxiliary consists of $N_{\text{aux}}$ sites with the same local Hilbert space $d$ as the physical sites of the system,
i.e. $d=2$ for $S=\frac{1}{2}$ spins, resulting in rank-3 tensors of the form $\chi_{i-1}\times\chi_i\times d$.
The number of auxiliary sites dictates the maximum bond dimension at the edge of the physical system
as $\chi_\text{aux} = d^{N_{\text{aux}}}$ and, hence,
the maximum amount of entanglement between the auxiliary and the system
\footnote{
Using $N_\mathrm{aux}$ spins of dimension $d$ is equivalent to preparing 
a single degree of freedom with $d^N_\mathrm{aux}$. 
However, the former has practical advantage regarding the ease of implementation and the physical intuition about the degrees of freedom included in the bath. 
}.
We emphasize that the auxiliary sites are not coupled to the physical system by any physical exchange,
and therefore only the identity is applied to them during the imaginary time-evolution.
\par
We extend TPQ-MPS to two spatial dimensions by wrapping the lattice on a cylinder with a finite circumference
and wind the 1D MPS structure around, enumerating all the sites linearly (see Fig.~\ref{f1}(c)).
Cylinder tensor networks are fairly standard techniques nowadays,
involving various variants in the way of wrapping the lattice and subsequent enumeration schemes.
The precise way of wrapping the lattice can have physical implications;
The system, although gapless in the two-dimensional limit, maybe gapped if the gapless nodes are not
on allowed momenta lines in the Brillouin zone\cite{Yin-Chen2017}. 
There are choices of particular cylindrical geometry known to capture the gapless state of the KH model 
\cite{Gohlke2017,Gohlke2018B}, but are not used here. 
The choice of such cylinder is important for the ground state but not for the temperature 
we can reach in the present study. 
The enumeration scheme should, ideally, not alter the physical properties. 
However, in reality, it can influence the spatial distribution of correlations and entanglement 
in particular for relatively small bond dimensions\cite{Li2019}. 
We employ a helical enumeration scheme with $8\times 3\times 2$ 
(YC8$\times$3$\times$2, which has circumference $L_\text{circ}=6$ and is illustrated in Fig.~\ref{f1}(c)) 
and $8\times4\times2$ sites (YC8$\times$3$\times$2, $L_\text{circ}=8$) 
conforming to YC3-$1$ and YC4-$1$, respectively, using the convention in Ref. \cite{Yan2011}. 
Both schemes treat the $x$- and $z$-bond on equal footing, 
i.e. they are nearest neighbors in the 1D MPS structure, 
while the $y$-bonds turn into an exchange with range $2L_\text{circ}-1$ sites. 
This choice results in the smallest $\chi_{MPO}$ of the time-evolution unitary, 
while also reducing the number of nearest-neighbor bonds cut by a bipartition which, at sufficiently low $T$, 
enters the amount of entanglement entropy encoded in the TPQ-MPS. 
\par
The long-range interactions within the effective 1D model make the time-evolving block decimation scheme
\cite{Vidal2004, White2004, Daley2004} in Eq.~\eqref{eqn:TPQ-state} infeasible.
Instead, we rely on an MPO formulation of the time-evolution operator\cite{Zaletel2015}
\footnote{We note that time-dependent variational principle (TDVP)~\cite{Haegeman2011,Haegeman2016}
can be utilized as well.}.
Specifically, we discretise $\mathcal U(\beta/2) = \left[\mathcal U(d\tau)\right]^N$ with small imaginary time steps
$d\tau$ and represent $\mathcal{U}(d\tau)$ as MPO\cite{Zaletel2015}.
\footnote{The MPO representation of the imaginary time evolution is given as
$W^{\rm II}(d\tau)\equiv \mathcal{U} (d\tau)$, following Ref.\cite{Zaletel2015}.
Splitting $d\tau = \tau_1 + \tau_2$ with sufficiently chosen complex $\tau_1$ and $\tau_2$
such that $\mathcal U(d\tau) \approx W^{\rm II}(\tau_2) W^{\rm II}(\tau_1)$
reduces the error in $d\tau$ by one order.
}
After each MPO-MPS product, the MPS is compressed
using a variational scheme \cite{Schollwoeck2011} reducing $\chi$.
We use an upper bound for the maximum $\chi_\text{}=\{362,512,724,1024\}$
to limit the computational resources needed.
If the bound is not reached, small Schmidt values $\lambda_i$ are discarded provided either
of the two criteria are met:
(I) discard all $\lambda_i \le s_\text{trunc}$ %$\lambda_i \le 10^{-6}$
or (II) discard all $\lambda_i$ sufficing $\sum_{i} \lambda_i^2 < (s^\text{sum}_\text{trunc})^2$
beginning from the smallest $\lambda_i$.
\par
Further technical details are given as follows;
The initial random TPQ-MPS state $|\Psi_0\rangle$ is prepared by applying a sequence of random two-site unitary matrices
to a N\'eel-like product state in the $z$-basis, i.e. $| \cdots \uparrow \downarrow \cdots \rangle$, in a TEBD-like way. 
We prepare $N_\mathrm{samples} = 100$ independent random initial states using 25 TEBD-iterations and cap the bond dimension at $\chi_\text{ini} = 32$.
See Appendix A and Ref.~\cite{Iwaki2022} for further details regarding the random initial state. 
The imaginary-time step is chosen as $d\tau = 0.1$ and smaller for $\beta\leq 0.8$.
Truncation thresholds are set to $s_\text{trunc} = 10^{-6}$ and $s^\text{sum}_\text{trunc} = 10^{-5}$ unless stated otherwise.
Measurements are not independent concerning $\beta$, but are done at certain series of
fixed $\beta$ during the single run of imaginary-time evolution and the $N_\text{samples}$ averages are taken from a set of independent runs.
The TenPy library \cite{Hauschild2018} is used for all MPS-related numerical calculations.
%
%%%%%%%%%%%%%%%%%%%%%%%%%%%%%%%%%%%%%%%%%%%%%%%%%%%%%%%%%%%%%%%%%%%%%%%%%%%%%%%
%%% Numerical results
%%%%%%%%%%%%%%%%%%%%%%%%%%%%%%%%%%%%%%%%%%%%%%%%%%%%%%%%%%%%%%%%%%%%%%%%%%%%%%%
\section{Application to the Kitaev honeycomb model}
We employ TPQ-MPS to the Kitaev honeycomb (KH) model defined as~\cite{Kitaev2006}
\begin{equation}
    \mathcal{H} = K_x \sum_{{\langle i,j\rangle_x}} \sigma^x_i \sigma^x_j 
    		+ K_y \sum_{{\langle i,j\rangle_y}} \sigma^y_i \sigma^y_j 
    		+ K_z \sum_{{\langle i,j\rangle_z}} \sigma^z_i \sigma^z_j~, 
\label{eqn:KitaevHam}    
\end{equation}
where $\sigma^\gamma_i$ are Pauli operators $\sigma^x$, $\sigma^y$, and $\sigma^z$ acting on sites~$i$. 
The three sets of parallel bonds on the honeycomb lattice 
are labeled as $\gamma = \{x,y,z\}$ (see Fig.~\ref{f1}(c)).
The Kitaev interaction $K_\gamma$ couples a neighboring pair of spins 
$\langle i,j \rangle_\gamma$ along the $\gamma$-bond by an Ising-like exchange $\sigma_i^\gamma \sigma_j^\gamma$. 
The KH model features a gapless QSL ground state if $K_\alpha \le K_\beta + K_\gamma$ is satisfied for 
all permutations of the bond labels $\{x,y,z\}$. 
Otherwise, a gapped QSL is found which adiabatically connects to the Toric Code \cite{Kitaev2003}. 
Here, we focus on the case of $K_x = K_y = K_z = \frac{1}{3}$.
\par
The KH model features a double-peak structure in the specific heat, 
signalling crossovers and releasing an entropy of $\Delta S/N = \frac{1}{2} \ln 2$ each. 
The associated two energy scales are well known\cite{Nasu2015}: 
At the high-$T$ peak, $T_H/K \approx 0.5$, 
nearest-neighbor spin-spin correlations develop and 
the fractionalization into itinerant and localized Majorana fermions occurs. 
The latter contributes to the formation of fluxes at each hexagonal plaquette given as 
 $W_{\mathcal P} = \prod_{i \in \mathcal P} \sigma^{\gamma_{\mathcal P}(i)}_i$~,
where $\gamma_{\mathcal P}(i)=x,y,z$ is the label of bond connected to site $i$ while not being 
part of the plaquette ${\mathcal P}$. 
The fluxes give an extensive set of quantum numbers, $w_{\mathcal P} = \pm 1$, 
which are disordered at $T\lesssim T_H/K$. 
Below the low-$T$ peak, $T_L/K \approx 0.016$, the fluctuation of fluxes is suppressed and 
we eventually find $\langle W_{\mathcal P}\rangle \rightarrow 1$. 
They form the static $\mathbb{Z}_2$ lattice-gauge field, fixing half of the Hilbert space per unit cell. 
A local Hilbert space dimension of $\sqrt 2$ per site remains which is associated with itinerant-free Majorana fermions. 
Although the KH model at finite temperature is not exactly solvable, 
once $\mathbb Z_2$ bond variables constituting the $\mathbb Z_2$ gauge field are treated as classical degrees of freedom, 
a combination of classical Monte Carlo method with free (Majorana) fermion exact diagonalization (MC$+$FFED) 
provides a nearly exact calculation in a relatively large cluster, as performed by Nasu, {\it et.al}\cite{Nasu2015}. 
Whereas, its counterpart Eq.(\ref{eqn:KitaevHam}) is a quantum many-body Hamiltonian 
which is generically difficult to solve at finite temperatures 
straightforwardly by an unbiased quantum many-body calculation. 
Therefore, the model provides a good platform and benchmark for our approach. 
We would like to emphasize that our approach, unlike MC$+$FFED, 
is not custom tailored to the Kitaev model and can be applied to other quantum many-body Hamiltonian. 
\par
%*%*%*%*%*%*%*%*%*%*%*%*%*%*
\begin{figure}[tb!]
    \centering
    \includegraphics[width=\linewidth]{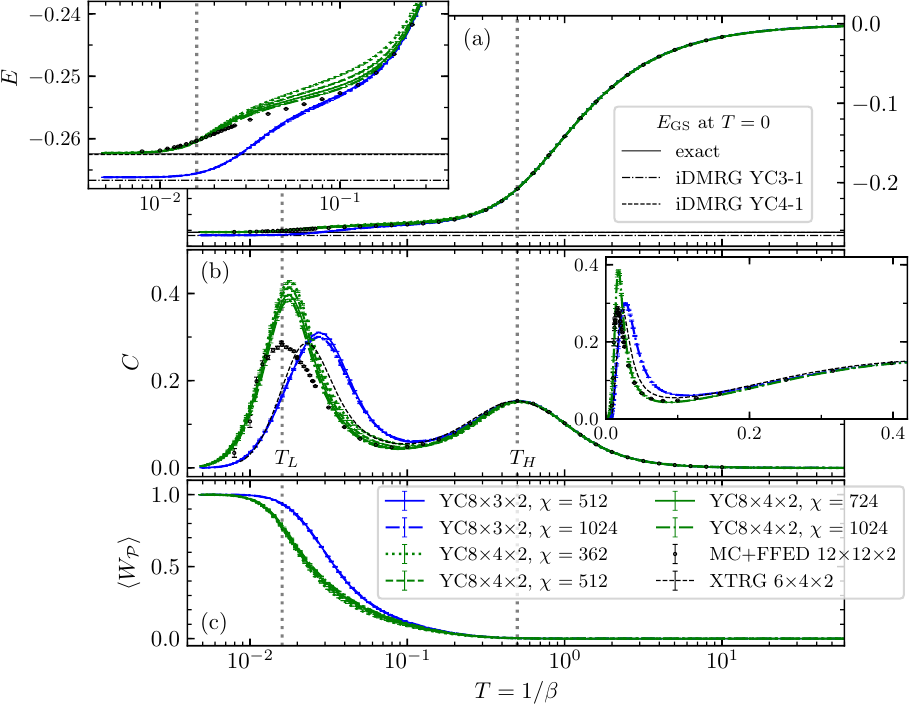}
    \caption{ 
	        Temperature ($T=\beta^{-1}$)-dependent (a) energy density $E$ and 
            (b) specific heat density exhibiting the double peak 
            obtained by TPQ-MPS on cluster YC8$\times$3$\times$2 and YC8$\times$4$\times$2
            and several upper bounds for $\chi_{}$. 
            Reference data (black dots) uses MC$+$FFED on $12\times12\times 2$ sites \cite{Nasu2015}. 
            YC8$\times$4$\times$2 exhibits a good quantitative agreement 
            with MC$+$FFED down to $T\sim 0.05$. 
            We attribute the difference in the position of $T_L$, in particular for YC8$\times$3$\times$2, 
            to the finite circumference geometry used here; 
            the ground state energies $E_{GS}$ of an infinitely long $L_\text{circ} =6$ cylinder 
            (YC3-1, dash-dotted horizontal line) obtained by iDMRG 
            has a ground state energy density lower than the bulk exact 
            one~\cite{Kitaev2006} (solid horizontal line) by about $\sim 0.01K$,
            yielding different crossover slopes in $E$ near $T_L$ 
            and a shifted peak. 
            The XTRG result using a $6\times 4\times 2$ cylinder is extracted from Fig.~4 in Ref.~\cite{Li2020} and shown for comparison.  
            The inset focuses on $T\leq 0.42$ using a linear scale for the temperature.
            (c) The evolution of the plaquette flux average, $\langle W_p\rangle$.
            The flux-free state at $T=0$ exhibits $\langle W_p \rangle =1$. 
      }
\label{f2}
\end{figure}
%*%*%*%*%*%*%*%*%*%*%*%*%*%*
%
Our TPQ-MPS data in Fig.~\ref{f2} 
exhibits a good qualitative agreement with the results obtained from MC$+$FFED~\cite{Nasu2015} 
on a $12\times12\times2$ cluster and XTRG using a $6\times4\times2$ geometry~\cite{Li2020}; 
The energy density\footnote{We are computing the energy density neglecting the left $N_l=L_\text{circ}$ and right $N_r=L_\text{circ}$ sites of the physical system to obtain a better estimate of the energy density in the bulk} $E$ rapidly decreases near $T_H$ resulting in a crossover peak in the specific heat $C$. 
A second step of energy reduction occurs near $T_L$. 
The two-step behavior is already present for small $\chi_\text{} = 362$
with well converged behaviour down to $T \sim 0.2$ including the high-$T$ peak in the specific heat. 
Whereas for $T \lesssim 0.2$, the finite-size and finite-$\chi$ effects inevitably influence the data; 
In Fig.~\ref{f2}(a) we display in two different lines 
the ground state energy obtained using iDMRG on an infinite cylinder with the same circumference $L_\text{circ}=6,8$ 
and helical boundary condition YC$3$-1 and YC$4$-1, respectively. 
The circumference seriously affect the numerically achieved ground state energies 
and consequently the specific heat which can be summarized as follows: 
(I) The cylinder with $L_\text{circ}=6$ features an enhanced reduction in energy upon cooling down
approaching the significantly lower ground state energy. 
The low-T peak in specific heat is of similar height to MC$+$FFED, while shifted to a two to three times higher temperature. 
(II) For $L_\text{circ}=8$ we obtain an evolution of the energy closer to MC$+$FFED, thus reducing the finite-size effect signicantly. 
For $T_L \gtrsim T \gtrsim 0.2$, however, TPQ-MPS overestimates $E$ compared to MC$+$FFED.
Here, increasing $\chi$ gradually reduces $E$ possibly approaching MC$+$FFED for sufficiently large $\chi$.
Near $T \sim T_L$ and below, the effect of finite $\chi$ ceases and the energy eventually approaches both MC$+$FFED as well as the ground state energy.
As a consequence of the overestimated energy density at intermediate $T$,
we obtain an enhanced slope of $E$ resulting in a higher peak in the specific heat. 
Again, increasing $\chi$ improves accuracy, reduces the height of the peak and results in a behaviour closer to MC$+$FFED.
The peak position is very similar to MC$+$FFED at any $\chi$. 
\par
The average of $\mathbb Z_2$ fluxes in Fig.~\ref{f2}(c) nicely marks the two peaks by 
an onset of nonzero value ($T_H$) and the inflection point ($T_L$), 
finally approaching $\langle W_{\mathcal P} \rangle \rightarrow 1$ at $T\rightarrow 0$ 
systematically for various $\chi$. 
\par
A recent XTRG calculation applied to the Kitaev model reports 
the lower-$T$-peak at $T_L\sim 0.023$ with the peak-height of $\sim 0.3$ 
using a $6\times 4\times 2$ cylinder~\cite{Li2020}. 
While the circumference is similar to our YC8$\times$4$\times$2, 
the XTRG work uses a slightly shorter cylinder, 
does not use helical boundary condition, and employs a different winding scheme.  
The quantitative agreement of XTRG with MC$+$FFED and TPQ-MPS is very good above $T \sim 0.1$ where finite-size effects become negligible.
At lower temperature, however, deviations become apparent (see Fig.~\ref{f2}): 
Our geometry YC8$\times$4$\times$2 with helical boundary condition exhibits a ground state energy close to the thermodynamic limit, whereas XTRG uses a different winding scheme, which influences the location of $T_L$. 
The comparison with our two geometries confirms that 
the size or shapes of the cylinder shifts the $T_L$ peak. 
The height of the peak in XTRG is similar to that of MC$+$FFED and YC8$\times$3$\times$2. 
Both TPQ-MPS and XTRG give reasonable results for the given finite size system, 
but the choice of the cylinder can easily influence the quantity of the data against 
the bulk data at $T\le 0.1$. 
\par
In this context, we like to remark that in many frustrated spin models, 
the specific heat at $T\lesssim 0.1$ naturally suffers large finite-size effect independent of the method employed. 
For example, in kagome-lattice Heisenberg antiferromagnet, 
specific choices of clusters sometimes yield unphysical peaks or features
not observed in other choices of cluster \cite{Sugiura2013,Shimokawa2016}
possibly obscuring the physical behaviour. 
%
%*%*%*%*%*%*%*%*%*%*%*%*%*%*%*%*%*
\begin{figure}
    \centering
    \includegraphics[scale=0.95]{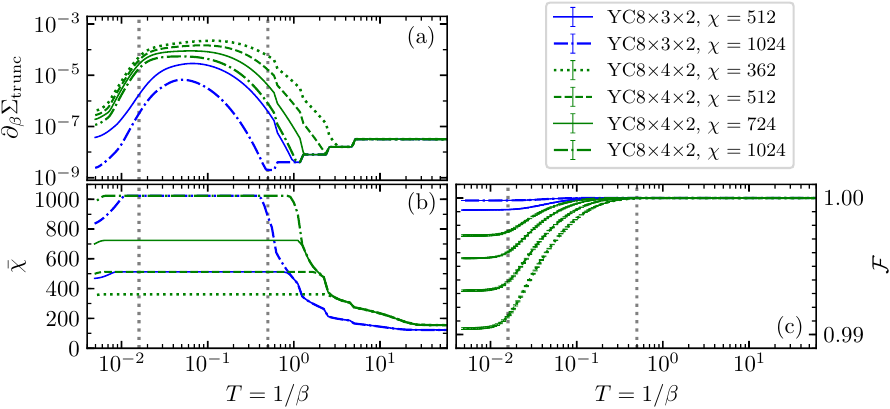}
    \caption{Evolution of the truncation error and bond dimension of the KH model:
    (a) Accumulated truncation error per unit of imaginary-time $\partial_\beta \Sigma_\text{trunc}$ 
   and bond dimension $\bar\chi$ averaged over the system,
    and (b) fidelity $\mathcal F$ of the evolved MPS 
before and after the truncation accumulated for all $\beta_j<\beta$. 
}
\label{f3}
\end{figure}
%*%*%*%*%*%*%*%*%*%*%*%*%*%*%*%*%*
%
%%%%%%%%%%%%%%%%%%%%%%%%%%%%%%%%%%%%%%%%%%%%%%%%%%%%%%%%%%%%%%%%%%%%%%%%%%%%%%%
%%% Truncation of the TPQ-MPS
%%%%%%%%%%%%%%%%%%%%%%%%%%%%%%%%%%%%%%%%%%%%%%%%%%%%%%%%%%%%%%%%%%%%%%%%%%%%%%%
\section{How truncation affects the TPQ-MPS state}
We now quantify the TPQ-MPS based on the error analysis during the run by focusing
on two quantities: 
The first one is the sum of all discarded Schmidt values ($\lambda_i$ for $i>i_0(\beta_j)$ which fulfills 
aforementioned I or II in $\S.$2) 
accumulated over a single imaginary-time evolution, 
\begin{equation}
\Sigma_\text{trunc} (\beta) = \sum_{\beta_j<\beta} \; \sum_{i>i_0(\beta_j)} \lambda_i^2(\beta_j)~. 
\end{equation}
The second one is the product of the fidelities of the state 
$|\Psi(\beta_{j})\rangle=W^{\rm II} (d\tau) | \Psi(\beta_{j-1}) \rangle$ (see Ref.[36]) and $|\tilde \Psi(\beta_{j})\rangle$
just before and after truncation, respectively, for all truncations down to the temperature $\beta^{-1}$, 
\begin{equation}
\mathcal F(\beta) = \prod_{\beta_j<\beta} (|\langle \tilde \Psi(\beta_{j})|\Psi(\beta_j)\rangle|^2)~, 
\end{equation}
which evaluates how we deviate from the non-truncated wave function at $\beta$. 
The amount of truncated Schmidt values per unit of imaginary time 
is given as $\partial_\beta \Sigma_\text{trunc}$.
In Figure~\ref{f3} we show the evolution of $\partial_\beta \Sigma_\text{trunc}$, of the average bond dimension $\bar \chi$,
and of $\mathcal{F}$. 
Upon cooling down, $\partial_\beta \Sigma_\text{trunc}$ remains below $10^{-7}$ until $\bar \chi$
reaches teh upper bound $\bar \chi \sim \chi$, which occurs near $T\sim 1$.
Larger $\chi$ (smaller system) generally lowers this threshold temperature. 
At these high temperatures, the evolution is very accurate reflected in a fidelity $\mathcal{F} \sim 1$.
Upon lowering the temperature, $\partial_\beta \Sigma_\text{trunc}$ 
increases and then reaches a plateau at $T_L \lesssim T \lesssim T_H$ 
with values $\partial_\beta \Sigma_\text{trunc} \sim 10^{-5}$ to $10^{-4}$ depending on $\chi$.
Here, $\mathcal F$ starts to depart gradually from $1$, which is more distinct for smaller $\chi$.
At $T\lesssim T_L$ the error 
$\partial_\beta \Sigma_\text{trunc}$ reduces again and $\mathcal F$ starts to flatten out. 
In particular for YC8$\times$3$\times$2, the drop in $\bar \chi$ is apparent,
indicating the reduction in the size of the Hilbert space needed to effectively encode the low-temperature state. 
\par
These observations suggest two effects of the truncation $\chi$;
For relatively small $\chi$ that is reached quickly, in particular at intermediate $T_L \lesssim T \lesssim T_H$, 
taking a larger $\chi$ lowers the energy towards the optimal value. 
This becomes evident upon inspection of $E$ in Fig.~\ref{f2}(a), 
whose accuracy improves for larger $\chi$ approaching the MC$+$FFED data. 
\par
The second effect concerns the states at high energy.
Let us expand the TPQ state constructed for the full Hilbert space for finite $N$. 
The system is split into a smaller part $A$ (with dimension $D_A$) and a bigger part $B$, 
which is Schmidt decomposed as 
\begin{equation}
      |\Psi_\beta \rangle = \sum_{n=1}^{D_A} \lambda_n |n_A\rangle |n_B\rangle
\end{equation}
to the orthogonal basis sets $\{|n_A\rangle\}$ and $\{|n_B\rangle\}$. 
The local part $A$ is thermalized and its density operator is approximated by the Gibbs state in $A$ as 
\begin{equation}
      \rho_A = \sum_{n=1}^{D_A} \lambda_n^2 |n_A\rangle\langle n_A| 
      \simeq \frac{e^{-\beta \mathcal{H}_A}}{Z_A} 
\end{equation}
where $\{|n_A\rangle\}$ is thought to be the energy eigenbasis 
of the subsystem's Hamiltonian $\mathcal{H}_A$. 
For its eigenvalues $\{E_n^A\}$, the Schmidt coefficient $\lambda_n$ is represented as 
$e^{-\beta E_n^A/2}/\sqrt{Z_A}$, and we find 
\begin{equation}
      |\Psi_\beta \rangle \simeq \sum_{n=1}^{D_A} \frac{e^{-\beta E_n^A/2}}{\sqrt{Z_A}}
      |n_A\rangle |n_B\rangle. 
\label{eq:abschmidt}
\end{equation}
Note here that $\{|n_B\rangle\}$ is left unknown. 
We finally truncate $|\Psi_\beta \rangle$ as $D_A\rightarrow \chi$ in Eq.(\ref{eq:abschmidt}), 
discarding the basis states with small weight. 
Specifically, information of $|n_A\rangle$ belonging to higher $E_n^A$ is lost. 
This explains the capability of TPQ-MPS to express qualitatively different quantum states from high to low temperatures;
The truncation of the MPS efficiently compresses the information needed to represent the thermal state in particular at low temperatures. 
\par
The above context of discarding high-temperature states---or high-energy states, respectively---efficiently, 
explains a particular feature of TPQ-MPS: the variance of physical quantities among different initial states 
becomes smaller by more than one order for lower temperature \cite{Iwaki2021,Iwaki2022}. 
This tendency is opposite to the usual random sampling methods including standard TPQ, 
or Monte Carlo methods, 
where the sampling error is by orders of magnitude larger in the lower temperature phase. 
We illustrate this point further by referring to the standard TPQ using the full Hilbert space 
for a limited system size $N$: 
It is shown in Ref.\cite{Sugiura2013} that the variance 
increases at low temperatures by the order of $\le e^{-S_N/T}$, and 
when the entropy $S_N$ of size $N$ is sufficiently large,
the increase is moderately suppressed. 
This fact supports the application of TPQ methods to highly frustrated quantum magnets 
including the present Kitaev model and kagome or related lattice models
\cite{Yamaji2016,Endo2018,Hotta2018,Rousochatzakis2019,Richter2022,Richter2023}. 
However, the idea of relying on the large entropy does not apply to TPQ-MPS: 
in the first proposal of TPQ-MPS in Ref.~\cite{Iwaki2021} some of the authors have shown 
that even for non-frustrated systems, the sample variance becomes smaller at lower temperature 
contrary to the prospect from TPQ. 
This is intuitively because MPS provides a good description of a quantum many body state at zero temperature. Our Eq.~(\ref{eq:abschmidt}) and the related discussions 
support this observation irrespective of the choice of spatial dimensions, 
and suggest good applicability of the present 2D TPQ-MPS to non-frustrated models.  
%
%%%%%%%%%%%%%%%%%%%%%%%%%%%%%%%%%%%%%%%%%%%%%%%%%%%%%%%%%%%%%%%%%%%%%%%%%%%%%%%
%%% Conclusion
%%%%%%%%%%%%%%%%%%%%%%%%%%%%%%%%%%%%%%%%%%%%%%%%%%%%%%%%%%%%%%%%%%%%%%%%%%%%%%%
\section{Conclusion}
To summarize, the TPQ-MPS is applied to 2D by wrapping the MPS train into cylinders.
The two peaks in the specific heat in the Kitaev honeycomb lattice signaling the fractionalization of spins into Majorana fermions 
and fixing the $\mathbb Z_2$ gauge flux are both reproduced.
While finite-size effects appear at $T \lesssim 0.1$ as is common with other methods,
finite-$\chi$ affects the MPS-TPQ only at intermediate temperatures $T\sim 0.1$
and is less of a concern at very low temperatures $T \lesssim 0.01$. 
This fact is in sharp contrast to other random sampling methods including the original TPQ method 
using the full Hilbert space.
Here, the truncation process of TPQ-MPS efficiently discards the higher-temperature information
explaining why it can track a nearly pure thermal state with its volume-law entanglement--equivalent to the thermal entropy--across a wide range of temperatures. 
This allows the state starting from random at high temperature (initial state) 
to gradually reach the qualitatively different long-range entangled topological ordered ground state.
The application to non-frustrated model is expected to be promising because, unlike for the original TPQ, a high entropy density is not required at low temperature to attain a reasonable accuracy at a moderate numerical cost. 
%
%
%%%%%%%%%%%%%%%%%%%%%%%%%%%%%%%%%%%%%%%%%%%%%%%%%%%%%%%%%%%%%%%%%%%%%%%%%%%%%%%
%%% Acknowledgements
%%%%%%%%%%%%%%%%%%%%%%%%%%%%%%%%%%%%%%%%%%%%%%%%%%%%%%%%%%%%%%%%%%%%%%%%%%%%%%%
\section*{Acknowledgements}
We thank J. Nasu for providing us with reference data. 
We acknowledge the use of computational resources of the supercomputer Fugaku provided by the RIKEN AICS through the HPCI System Research Project (Project ID: hp210321)
and of the Scientific Computing section of the Research Support Division at the Okinawa Institute of Science and Technology Graduate University (OIST).
M.G.~acknowledges support by the Theory of Quantum Matter Unit at OIST.
%
% TODO: include author contributions
%\paragraph{Author contributions}
%This is optional. If desired, contributions should be succinctly described in a single short paragraph, using author initials.

% TODO: include funding information
\paragraph{Funding information}
This work was supported by a Grant-in-Aid for Transformative Research Areas "The Natural Laws of Extreme Universe---
A New Paradigm for Spacetime and Matter from Quantum Information" (No. 21H05191) 
and JSPS KAKENHI (Grants No. JP21K03440 and JP22K14008).
A.I. was supported by JSPS Research Fellowship (Grant No. 21J21992). 

%Authors are required to provide funding information, including relevant agencies and grant numbers with linked author's initials. Correctly-provided data will be linked to funders listed in the \href{https://www.crossref.org/services/funder-registry/}{\sf Fundref registry}.
%
%%%%%%%%%%%%%%%%%%%%%%%%%%%%%%%%%%%%%%%%%%%%%%%%%%%%%%%%%%%%%%%%%%%%%%%%%%%%%%%
%%% Appendix
%%%%%%%%%%%%%%%%%%%%%%%%%%%%%%%%%%%%%%%%%%%%%%%%%%%%%%%%%%%%%%%%%%%%%%%%%%%%%%%
\begin{appendix}
\section{Random sampling average}
TPQ-MPS is a random sampling method using the MPS representation of the quantum many-body wave function.
Since the quality of the MPS state relies its bond dimension $\chi$ practically accessible in the computation,
a smaller $\chi$ would require a larger number of independent runs to be averaged over. 
This number is generally by orders of magnitude smaller than with METTS 
when applying them to the same system. 
\par
Let us first highlight the difference from METTS,
which is a moderately mixed quantum thermal state. 
METTS starts from a (classical) product state with $\chi=1$,
and $\chi$ grows upon imaginary time evolution.
Accordingly, the entanglement stored is moderate and does not suffice the amount of entropy of the thermal state: 
This is compensated by taking an average over samples and constructing a mixed state. 
In METTS, a Markov chain scheme is employed to increase the efficiency of the sampling,
while the sample number typically amounts to order-10$^2$.
Their $\chi$ can be kept smaller by increasing the sample average. 
TPQ-MPS does not rely much on the sample average as by storing the entanglement as maximally as possible in a single pure TPQ-MPS form, 
which usually requires large $\chi$. 
However, by relying on the auxiliary degrees of freedom on the edges,
the entanglement entropy distribution becomes nearly flat 
(without auxiliaries, $\chi$ at the edge becomes very small),
and this allows us to store the entanglement maximally for a given $\chi$. 
There exists a trade-off between the size of $\chi$ and of the sample average,
but each method has a specific control parameter to keep them 
in the practically reasonable range. 
A more detailed comparison with METTS is given in Ref.~\cite{Iwaki2022}. 
\par
One may anticipate that the present 2D Kitaev honeycomb (KH) model might require more samples
than for 1D systems\cite{Iwaki2021}.
As illustrated in Fig.~\ref{f1} in the main text, the higher the purity is,
fewer samples $N_{\rm samp}$ are needed to safely reproduce the thermal quantum state.
In TPQ-MPS, the $N_{\rm sample}$-independent runs are performed starting from
the independent initial random MPS, yielding a set of
{\it unnormalized} states over different $\beta_j$ for each, $\{|\Psi^{(l)}(\beta_j)\rangle\}_{l=1}^{N_{\rm sample}}$.
The random average of physical quantities $\mathcal O$ is taken as
\begin{equation}
\langle \mathcal{O} \rangle=
\frac{\sum_{l=1}^{N_{\rm sample}} \langle \Psi^{(l)}(\beta_j) | \mathcal{O} |\Psi^{(l)}(\beta_j)\rangle}
{\sum_{l=1}^{N_{\rm sample}} \langle \Psi^{(l)}(\beta_j)|\Psi^{(l)} (\beta_j)\rangle }.
\label{eq:randomav}
\end{equation}
Here, the summations of samples are taken independently between the numerator and denominator,
since the partition function is given by the denominator
$Z=\sum_{l=1}^M \langle \Psi^{(l)}(\beta_j) |\Psi^{(l)}(\beta_j)\rangle$
(the reasoning for why the average taken by the normalized $|\Psi^{(l)}(\beta_j)\rangle$
does not provide the correct sampling average is analytically shown in Ref.~\cite{Iwaki2022}).
\par
The number $N_{\rm sample}$ required can be measured using a quantity called
normalized fluctuation of partition function (NFPF),
\begin{equation}
\delta z^2= \frac{{\rm Var}( \langle \Psi(\beta_j) |\Psi(\beta_j)\rangle ) }{
\left(\overline{\langle \Psi(\beta_j) |\Psi(\beta_j)\rangle}\right)^2},
\label{eq:nfpf}
\end{equation}
where $\overline{\cdots}$ is the random average.
It is shown that the purity of the thermal state scales with $\delta z^2$
and the larger $\delta z^2$ means that the obtained state varies much with a sample.
In fact, we showed that the number of samples needed, $N_{\rm sample}$, to obtain the same quality
of Eq.(\ref{eq:randomav}) increases proportionally to $\delta z^2$ \cite{Iwaki2022}.
\par
%*%*%*%*%*%*%*%*%*%*%*%*%*%*%*%*%*
\begin{figure}[tb]
    \centering
    \includegraphics[width=\linewidth]{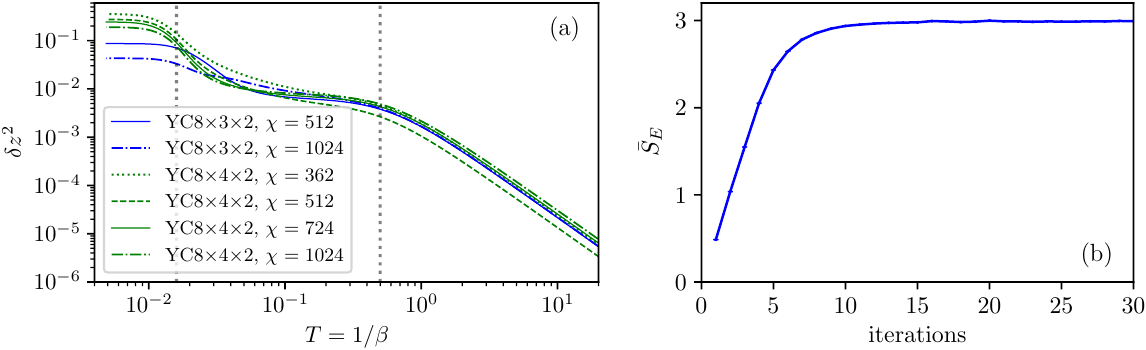}
    \caption{(a) Normalized fluctuation of partition function (NFPF) $\delta z^2$ as a function of $T$
    for the data calculated in Fig.~2 for the KH model with different $\chi$.
    The number of samples required to obtain the same quality data scales linearly with $\delta z^2$. 
    (b) Evolution of the entanglement entropy of bipartition as a function of iterations.
    Each iteration consists of applying two-site random unitary matrices on even and odd bonds similar to the TEBD algorithm.
    $\bar{S_E}$ is obtained by taking the average of $S_E$ over bipartitions between any of the physical sites.
    An additional sample averaging is done over 10 samples to obtain error bars indicating the variance. However, error bars are typically smaller than the marker size.
}
\label{fadd}
\end{figure}
%*%*%*%*%*%*%*%*%*%*%*%*%*%*%*%*%*
The results of $\delta z^2$ for the present calculation on the KH model are given in Fig.~\ref{fadd}(a)
for a set of data given in Fig.~\ref{f2}.
The largest $\delta z^2$ at low-$T$ ranges at $10^{-1}-10^1$, which is comparable to the value
for the 1D Heisenberg model\cite{Iwaki2022} which used $N_{\rm sample}=100$.
Based on this comparison, we also adopt $N_{\rm sample}=100$ for the 2D case.
The present calculation shows that the 2D TPQ-MPS is as capable as the 1D case despite the consensus
that the calculations in 2D are much more difficult than in 1D.
\par
The plateau of $\partial_\beta \Sigma_\text{trunc}$ observed in Fig.~\ref{f3} agrees with the plateau of $\delta  z^2$,
and as in Fig.~\ref{f3}, $\chi$ dependence appears at $T\lesssim 10^0$,
showing that $\delta z^2$ is indeed a good measure to qualify the quantum state.
We find a suppression of $\delta z^2$ by a log scale in terms of $\chi$,
indicating the high capability of TPQ-MPS to store the information required for
a wide range of temperatures exhibiting different natures.
\par
As mentioned above, TPQ-MPS stores substantial amount of entropy in the starting point of the imaginary time evolution, 
which is contrary to METTS. 
Therefore, the quality of the initial random state is important to have high purity and smaller $\delta z^2$, respectively. 
When generating the initial random TPQ-MPS state,
we use a TEBD-like algorithm with alternating application of random two-site unitary matrices.
Although not relevant to our model, an advantage of this method is the possibility of utilizing charge or $S^z$ conservation,
e.g. when studying a $U(1)$ symmetric model. 
The iteration number, we use 25 iterations, is determined by ensuring a saturated entanglement entropy. 
Figure~\ref{fadd}(b) shows the evolution of entanglement entropy of bipartition $\bar{S_E}$, averaged over the physical sites, as a function of iterations. 
While the bond dimension of the state generally doubles after each iterations, 
$\bar{S_E}$ is not fully saturated after $N=5$ iterations, which is required to reach $\chi_\text{RMPS} = 32$. 
Instead, saturation is reached at around $N\sim 10$ iterations. 
Our choice of $N=25$ iterations is well within the saturated regime. 
%%%%%%%%%%%%%%%%%%%%%%%%%%%% Bibliography %%%%%%%%%%%%%%%%%%%%%%%%%%%%%%%%%%%%%

\end{appendix}
% \bibliographystyle{SciPost_bibstyle} % Include this style file here only if you are not using our template
\bibliography{references}
\nolinenumbers
\end{document}